\documentclass[structabstract]{aa}  
\usepackage{graphicx}
\usepackage{amsmath}
\usepackage{natbib}
\newcommand{\degree}{\ensuremath{^{\circ}}}
\newcommand{\modif}{}
\begin{document}
   \title{ASTEP South: \\ An Antarctic Search for Transiting ExoPlanets \\ around the celestial South pole}


   \author{Crouzet N. \inst{1}, Guillot T. \inst{1}, Agabi A. \inst{2}, Rivet J.-P. \inst{1}, Bondoux E. \inst{2,5}, Challita Z. \inst{2,5}, Fante\"\i-Caujolle Y. \inst{2}, Fressin F. \inst{3}, M\'ekarnia D. \inst{2}, Schmider F.-X. \inst{2}, Valbousquet F. \inst{4}, Blazit A. \inst{2}, Bonhomme S. \inst{1}, Abe L. \inst{2}, Daban J.-B. \inst{2}, Gouvret C. \inst{2}, Fruth T. \inst{6}, Rauer H. \inst{6,7}, Erikson A. \inst{6}, Barbieri M. \inst{8}, Aigrain S. \inst{9}, Pont F.\inst{9}}

   \institute{Universit\'e de Nice Sophia Antipolis, CNRS UMR 6202, Observatoire de la C\^ote d'Azur, 06304 Nice Cedex 4, France\\
     \email{Nicolas.Crouzet@oca.eu}
     \and
     Universit\'e de Nice Sophia Antipolis, CNRS UMR 6525, Observatoire de la C\^ote d'Azur, 06108 Nice Cedex 2, France
     \and
     Harvard-Smithsonian Center for Astrophysics, 60 Garden Street, Cambridge, MA 02138, United States
     \and
     Optique et Vision, 6 bis avenue de l'Est\'erel, BP 69, 06162 Juan-Les-Pins, France
     \and
     Concordia Station, Dome~C, Antarctica
     \and
     DLR Institute for Planetary Research, Rutherfordstrasse 2, 12489 Berlin, Germany
     \and
     Center for Astronomy and Astrophysics, TU Berlin, Hardenbergstr. 36, 10623 Berlin, Germany
     \and
     Universit\`a di Padova, Dipartimento di Astronomia, vicolo dell'Osservatorio 5, 35122 Padova, Italia
        \and
School of Physics, University of Exeter, Stocker Road, Exeter EX4 4QL, United Kingdom}

   \date{December 2, 2009}
   

   \abstract
   {The Concordia base in Dome~C, Antarctica, is an extremely promising site for photometric astronomy due to the 3-month long night during the Antarctic winter, favorable weather conditions, and low scintillation.}
   {The ASTEP project (Antarctic Search for Transiting ExoPlanets) is a pilot project to discover transiting planets and understand the limits of visible photometry from the Concordia site.}
   {ASTEP South is the first phase of the ASTEP project. The instrument is a fixed 10~cm refractor with a 4kx4k CCD camera in a thermalized box, pointing continuously a $3.88\times 3.88\degree$$^2$ field of view centered on the celestial South pole. We describe the project and report results of a preliminary data analysis.}
   {ASTEP South became fully functional in June 2008 and obtained 1592 hours of data during the 2008 Antarctic winter. The data are of good quality but the analysis has to account for changes in the PSF (Point Spread Function) due to rapid ground seeing variations and instrumental effects. The pointing direction is stable within 10 arcseconds on a daily timescale and drifts by only 34 arcseconds in 50 days. A truly continuous photometry of bright stars is possible in June (the noon sky background peaks at a magnitude R $\approx 15\,\rm arcsec^{-2}$ on June 22), but becomes challenging in July (the noon sky background magnitude is R $\approx 12.5\,\rm arcsec^{-2}$ on July 20). The weather conditions are estimated from the number of stars detected in the field. For the 2008 winter, the statistics are between 56.3~\% and 68.4~\% of excellent weather, 17.9~\% to 30~\% of veiled weather (when the probable presence of thin clouds implies a lower number of detected stars) and 13.7~\% of bad weather. Using these results in a probabilistic analysis of transit detection, we show that the detection efficiency of transiting exoplanets in one given field is improved at Dome~C compared to a temperate site such as La Silla. For example we estimate that a year-long campaign of 10~cm refractor could reach an efficiency of 69~\% at Dome~C versus 45~\% at La Silla for detecting 2-day period giant planets around target stars from magnitude 10 to 15. The detection efficiency decreases for planets with longer orbital periods, but in relative sense it is even more favorable to Dome~C.}
   {This shows the high potential of Dome~C for photometry and future planet discoveries.}

   
   \authorrunning{Crouzet et al.}
   \titlerunning{ASTEP South}
   
   \maketitle
%

\section{Introduction}

Dome~C offers exceptional conditions for astronomy thanks to a 3-month continuous night during the Antarctic winter and a very dry atmosphere. Dome~C is located at $\rm{75\degree 06'S - 123\degree 21'E}$ at an altitude of 3233 meters on a summit of the high Antarctic plateau, 1100 km away from the coast. After a pioneering summer expedition in 1995, the site testing for astronomy begun in the early 2000's. It revealed a very clear sky, an exceptional seeing and very low wind-speeds \citep{Aristidi2003,Aristidi2005a,Lawrence2004a,Ashley2005a,Geissler2006b}. The French-Italian base Concordia was constructed at Dome~C from 1999 to 2005 to hold various science experiments. Summer time astronomy experiments have been carried out \citep[e.g.][]{Guerri2007}. The study of Dome~C for astronomy during night-time has considerably expanded since the first winter-over at Concordia in 2005. The winter site testing has shown an excellent seeing above a thin boundary layer \citep{Agabi2006,Trinquet2008,Aristidi2009}, a very low scintillation \citep{Kenyon2006a} and a high duty cycle \citep{Mosser2007a}. Low sky brightness and extinction are also expected \citep{Kenyon2006b}.

Time-series observations such as those implied by the detection of transiting exoplanets should benefit from these atmospherical conditions and the good phase coverage. This could potentially greatly improve the photometric precision when compared to other temperate sites \citep{Pont2005a}. A first photometric instrument, PAIX \citep{Chadid2007}, was installed at Concordia in December 2006. A lightcurve of the RR Lyrae variable star Sara over 16 nights in August 2007 is presented in \citet{Chadid2008}, and results of the whole campaign from June to August 2007 have been submitted. The sIRAIT instrument also obtained lightcurves over 10 days on the stars V841 Cen and V1034 Cen \citep{Briguglio2009,Strassmeier2008}.  

The ASTEP project (Antarctic Search for Transiting ExoPlanets) aims at determining the quality of Dome~C as a site for future photometric surveys and to detect transiting planets \citep{Fressin2005a}. The main instrument is a 40~cm Newton telescope entirely designed and built to perform high precision photometry from Dome~C. The observations will start in winter 2010. A first instrument already on site, ASTEP South, has observed during the 2008 and 2009 winters.

We present here the ASTEP South project and results from the preliminary analysis of the 2008 campaign. We first describe the instrument, the observation strategy and the field of view. Section~\ref{sec:data} discusses the main features obtained when running this simple instrument from Dome~C: influence of the Sun and the Moon, PSF and pointing variations, as well as temperature effects. In section~\ref{sec:duty cycle} we detail our duty cycle and infer the weather statistics at Dome~C for the 2008 winter. These results are combined to a probabilistic analysis to infer the potential of ASTEP South for planet detection and to evaluate Dome~C as a site for future planet discoveries.

\section{Instrumental setup}

\subsection{The instrument}

\begin{figure}[ht]
\centering
\resizebox{8.8cm}{!}{\includegraphics{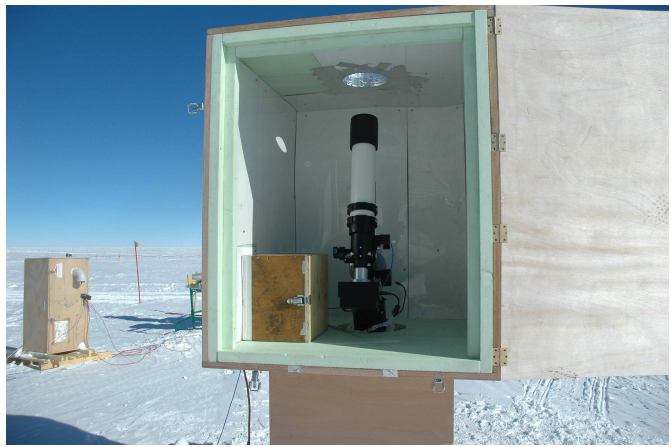}}
\caption{ASTEP South at Dome~C, Antarctica, January 2008.}
\label{fig:ASTEPSouthDomeC}
\end{figure}

ASTEP South consists of a 10~cm refractor, a front-illuminated 4096x4096 pixels CCD camera, and a simple mount in a thermalized enclosure. The refractor is a commercial TeleVue NP101 and the camera is a ProLines series by Finger Lake \modif{Instrumentation} equipped with a KAF-16801E CCD by Kodak. For the choice of the camera see \citet{Crouzet2007}. Its quantum efficiency peaks at 63~\% at 660~nm and is above 50~\% from 550 to 720~nm. The pixel size is 9 $\mu$m and the total CCD size is 3.7~cm. The pixel response non-uniformity is around 0.5~\%. Pixels are coded on 16 bits, implying a dynamic range of 65535~ADU. The gain is 2 e-/ADU. A filter whose transmission starts at 600~nm is placed before the camera to eliminate blue light. Given the CCD quantum efficiency, the overall transmission (600 to 900\,nm) is equivalent to that of a large R band. We use a GM 8 equatorial mount from Losmandy. A thermalized enclosure is used to avoid temperature fluctuations. The sides of this enclosure are made with wood and polystyrene. A double glass window reduces temperature variations and its accompanying turbulence on the optical path. Windows are fixed together by a teflon part and separated by a 3~mm space filled with nitrogen to avoid vapour mist. The enclosure is thermalized to $-$20\degree C and fans are used for air circulation. The ASTEP South instrument is shown at Dome~C in figure~\ref{fig:ASTEPSouthDomeC}.

In order to characterize the quality of Dome~C for photometric observations, we have to avoid as much as possible instrumental noises and in particular jitter noise, leading to a new observation strategy: the instrument is completely fixed and points towards the celestial South pole continuously. This allows also a low and constant airmass. The observed field of view is $3.88\times 3.88\degree$$^2$, leading to a pixel size of 3.41 arcsec on the sky. This field contains around 8000 stars up to magnitude R~=~15. This observation setup leads to stars moving on the CCD from frame to frame and to a widening of the PSF (Point Spread Function) in one direction, depending on the exposure time.

Test observations were made at the Calern site (Observatoire de la C\^ote d'Azur) observing the celestial North \modif{pole}, in order to choose the exposure time and the PSF size. A 30 second exposure time and a 2 pixel PSF FWHM (Full Width Half Maximum) lead to only 2 saturated stars and a limit magnitude around 14 (from Dome~C the limit magnitude is increased to 15). An analysis of the celestial South pole field from the Guide Star Catalog \modif{GSC2.2} with these parameters taking into account the rotation of the star during each exposure leads to less than 10~\% of blended stars. Therefore we adopted these parameters.

Software programs were developed by our team to control the camera, to run the acquisitions and to transfer and save the data. The instrument was set up at the Concordia base in January 2008.



\subsection{The South pole field}
\label{sec:The South pole field}

\begin{figure}[ht]
\centering
\includegraphics[width=9cm]{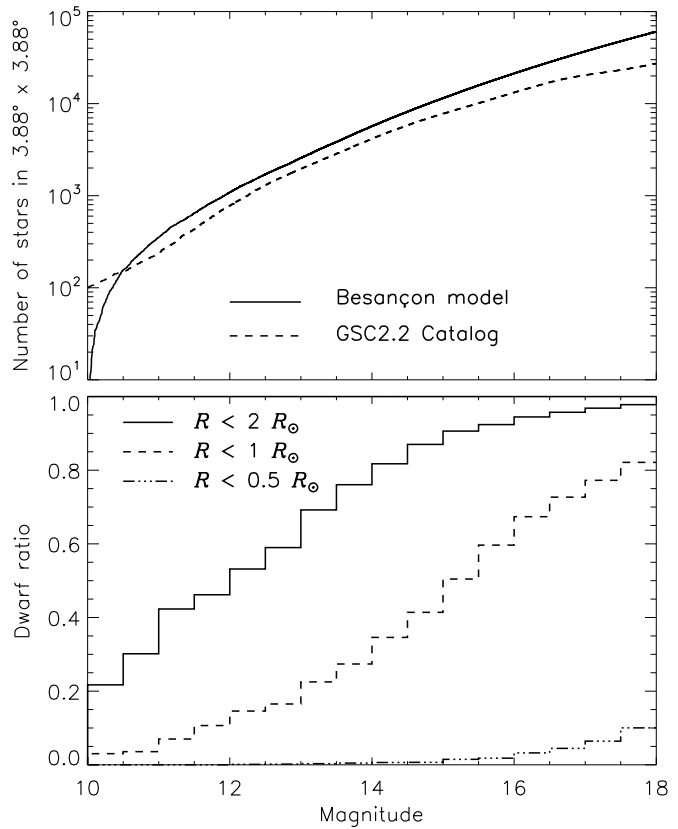}
\caption{{\it Top panel\/}: Cumulative distribution of the number of stars in the South pole field as a function of their magnitude in the R band. The plain line shows results from the Besan\c{c}on model. The dashed line indicates output from the GSC2.2 catalog. {\it Bottom panel}: Ratio of dwarf stars with selected radii (less than 2, 1 and 0.5\,$R_\odot$, respectively, as labeled) to the total number of stars in the South pole field as a function of R magnitude. }
\label{fig:mag-distrib}
\end{figure}

The distribution of stars in our field of view is shown in figure~\ref{fig:mag-distrib}. From the GSC2.2 catalog, we find nearly 8000 stars up to our limit magnitude of 15. We also simulate stellar populations in a field of 3.88$\times$3.88\degree$^2$ centered on the celestial South pole using the Besan\c{c}on model of the Galaxy\footnote{\tt http://bison.obs-besancon.fr/modele/} \citep{Robin2003} for R-band magnitudes between 10 and 18 to calculate the dwarf ratio in the field. The comparison shows that the Besan\c{c}on model overestimates the number of stars in the field by a factor $\sim 2$. However, we believe that the ratio of dwarfs to the total number of stars is, by construction of the model, better estimated. The bottom panel of fig.~\ref{fig:mag-distrib} shows that most of the stars brighter than magnitude R~=~12 are giants (or more accurately larger than twice our Sun).

Table~\ref{tab:nbstars} details the number of stars per magnitude range; the total number of stars is obtained from the GSC2.2 catalog and the number of dwarfs is estimated using the relative fractions from the Besan\c{c}on model. From magnitude 10 to 15 we have 73.6~\% of dwarf stars with radius $R < 2R_\odot$. This ratio is higher than in other typical fields used in the search for transiting planets such as Carina. Based on CoRoTlux simulations \citep{Fressin2007a}, we expect that about one F, G, K dwarf in 1100 to 1600 should harbor a transiting giant exoplanet. The South pole field observed by ASTEP South is thus, in principle, populated enough for the detection of transiting planets \citep[see also][]{Crouzet2009}. We will come back to a realistic estimate of the number of detectable exoplanets in section~\ref{sec:Planet detection probability}.

The advantages of the South pole field are hence of course a continuous airmass, a high ratio of dwarfs to giant stars and a very low contamination by background stars. On the other hand, the field is less dense than regions closer to the galactic plane, so that the actual number of transiting planets in the field is smaller. 



\begin{table}
\caption{Number of stars in the 3.88$\times$3.88\degree$^2$ celestial South pole field.}
\label{tab:nbstars}
\begin{center}
\begin{tabular}{rrrrrr}
\hline\hline
Magnitude & 10-11 & 11-12 & 12-13 & 13-14 & 14-15 \\ \hline
Total & 133 & 545 & 1171 & 2190 & 3608 \\
$R < 2R_\odot$: & 35 & 243 & 662 & 1605 & 3057 \\
$R < 1R_\odot$: & 4 & 50 & 184 & 556 & 1388 \\
$R < 0.5R_\odot$: & 0 & 0 & 2 & 9 & 24 \\ \hline\hline
\end{tabular}
\end{center}
\end{table}

\subsection{Temperature conditions}

The instrument was set up during the Dome~C 2008 summer campaign. The external temperature varied at this time between $-$20 and $-$30\degree C. It was let outside without thermal control until the observations started at the end of April. In winter the external temperature varies between $-$50 and $-$80\degree C. During the observations, the thermalized box is set to a temperature of $-$20\degree C and the CCD to $-$35\degree C. \modif{Because of self-heating,} the electronics of the camera is around $+$5\degree C with some variations (see \ref{sec:Camera temperature variations}).

\section{Preliminary data analysis}
\label{sec:data}

ASTEP South generates around 60 gigabytes of data per day. Since internet facilities at Dome~C are limited to a low stream connection only few hours a day, a whole data transfer is impossible. Data are stored in external hard disks and repatriated at the end of the winter-over, leading to at least a 6 month delay between the observations and a full data analysis. We thus developed a software program for on-site preliminary data analysis, in order to have a day-to-day feedback of the observations. We detail here the results of this preliminary analysis.

\subsection{Preliminary data analysis software program}

We developed a software program running on the data at Concordia. For each image of a given day the mean intensity is computed. We then process only the $1000\times 1000$ pixel central part of the frame ($0.95\times 0.95\degree{^2}$) for faster calculations. First, a point source identifier gives the number of detected stars and their location on the CCD. The 200 brightest stars are matched to the GSC 2.2 catalog using a home-made algorithm, in order to identify the South pole on the CCD. The 30 brightest stars are fitted with a gaussian to derive the PSF size. Last, basic aperture photometry is performed for a set of 10 stars without any image calibration. The identification of point sources, the gaussian fit and the aperture photometry use an IDL version of DAOPHOT \citep{Stetson1987}. A point source is considered as a star if its flux is 5 times larger then the sky noise. Aperture \modif{photometry} is made with large apertures of diameter 12 and 20 pixels, allowing to get all the flux for bright stars. Although these large aperture are not adapted to faint stars, the low crowding in our field allows to get reasonable lightcurves. Of course this will be optimized during the complete analysis of data. The camera and CCD temperature are also recorded. A small size binary file with these results is sent everyday by email. Plots shown in the following are in UTC time as recorded by the software program (local time at Dome~C is UTC~$+$8).

\subsection{Magnitude calibration}
\label{sec:calibration}

In order to convert ADU into magnitudes, we perform a preliminary magnitude calibration: we measure the flux of the stars on a typical image taken under dark sky and convert them into instrumental magnitudes. We then compare these magnitudes to the ones from the GSC2.2 catalog and obtain the so-called zero point. The image used is a raw image, but the local background including bias is \modif{subtracted} when calculating the flux of each stars. 

We estimate that the error on these magnitudes should be $\pm 0.3$ magnitudes or less. First a comparison of the result for all the stars in a given image to that obtained with only the stars in the $1000\times 1000$ pixel central part yields a 0.2 magnitude difference. We estimate that the absence of a flat-field procedure is responsible for that difference and that its impact on our inferred sky brightness magnitude should be smaller. Second, while one may estimate that the GSC2.2 errors on the magnitudes of individual stars can be as large as 0.5, the large number of stars ($\sim 7000$) implies that the mean error should be quite smaller. A 0.3 error on the inferred magnitudes hence appears to be a conservative estimate. 

In what follows, we will use this ADU to magnitude conversion only for the noon and full-moon sky brightness, not for the dark sky. This is because our preliminary analysis is based on data processed on the fly in Concordia which have not been de-biased. Variations in the bias level are of the order of 40~ADU. Given that uncertainty, we estimate that any measurement of magnitudes larger than 18 may have a bias error larger than 0.3 magnitudes and therefore refrain from mentioning those.  


A refined analysis of the full ASTEP South data with all available data is under way and will include an accurate de-biasing and magnitude calibration. 


\subsection{Influence of the Sun}

\begin{figure}[ht]
\centering
\includegraphics[width=8.5cm]{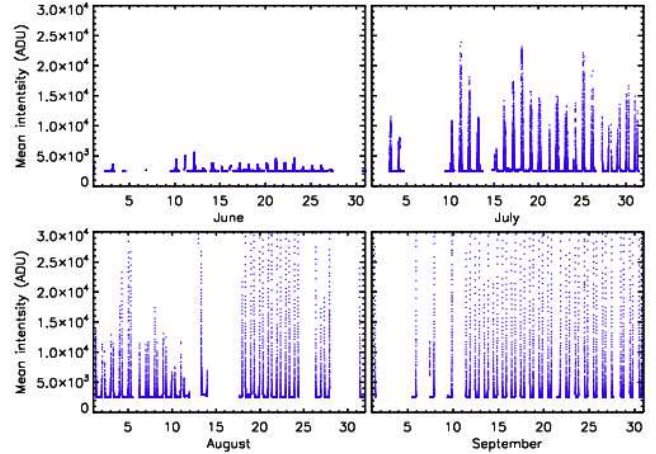}
\caption{Image intensity for all the winter. The bias level is around 2480~ADU and the scale is half of the CCD dynamics. Peaks around noon will affect the photometry.}
\label{fig:Intensmonth_sun}
\end{figure}

We first consider the influence of the Sun on the photometry. It is important to notice that although the Sun disappears below the horizon from May~4 to August~9, the sky background is always higher each day in the period around noon which is therefore less favorable for accurate photometric measurements. The minimum altitude of the Sun at noon occurs on June~21 and is $8.5\degree$ below the horizon. \modif{The height and width of the peak of intensity are the smallest around the winter solstice and increase before and after this date (figure~\ref{fig:Intensmonth_sun}). The increase is not linear but varies from one day to another, as also observed with the sIRAIT instrument \citep{Strassmeier2008}. We attempted to check whether this may be due to high altitude clouds but no correlation was found between the sky brightness and the quality of the night derived by studying the number of detected stars (see section~\ref{sec:duty cycle}).} 

\begin{figure}[ht]
\centering
\includegraphics[width=9cm]{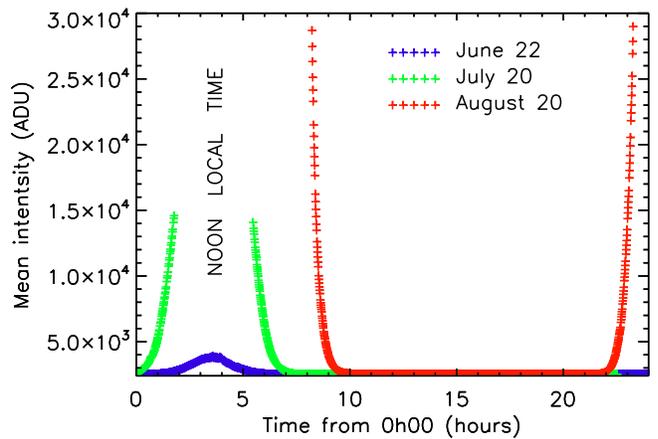}
\caption{Image intensity during 3 typical days in June, July and August. The bias level is around 2480~ADU and the scale is half of the CCD dynamics. Due to the Sun, the sky background increases around noon. The corresponding magnitude for June~22 is 15.3~mag\,/\,arcsec$^2$.}
\label{fig:Intensday}
\end{figure}

Figure~\ref{fig:Intensday} shows variations of the mean intensity as a function of time for 3 clear days: June~22, July~20 and August~20. On June, 21 the height is typically 1600~ADU and images are affected during 4 to 6 hours. From our calibration this corresponds to a magnitude of $15.3\,\rm arcsec^{-2}$ in the standard R band. The residual noise calculated from the actual number of photons received from the sky in an aperture of 20 pixels (corresponding to a radius equal to a FWHM of 2.5 pixels) is $4\times 10^{-3}$. For larger apertures the noise will be smaller. Therefore this effect will have a moderate impact on the photometry. In July the height grows to typically 20000~ADU, i.e. magnitude $12.6\,\rm arcsec^{-2}$, and a noticeable brightness increase lasts for 7 to 9 hours. In August, this brightness increase lasts 9 to 12 hours. 

\begin{figure}[ht]
\centering
\includegraphics[width=9cm]{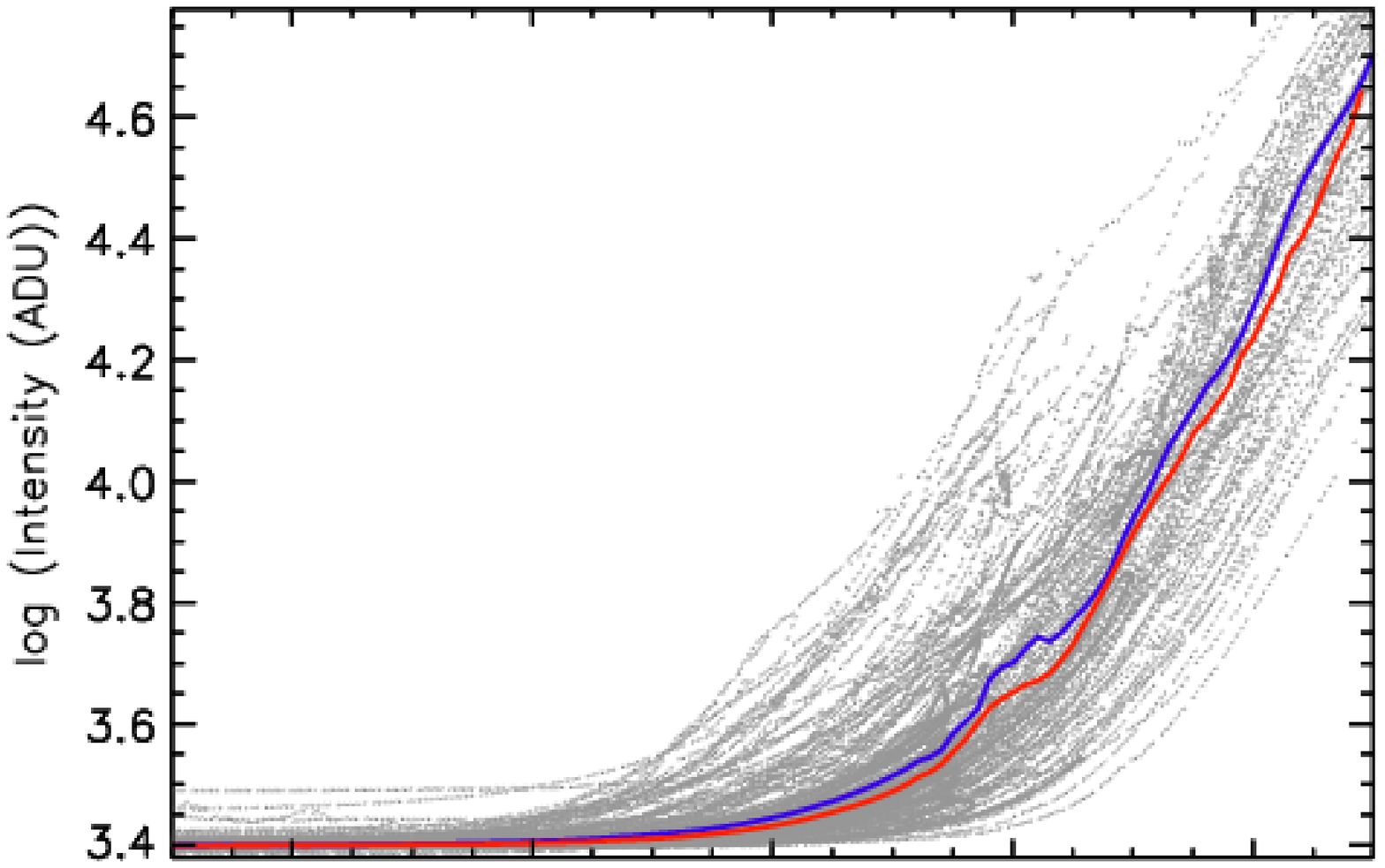}
\includegraphics[width=9cm]{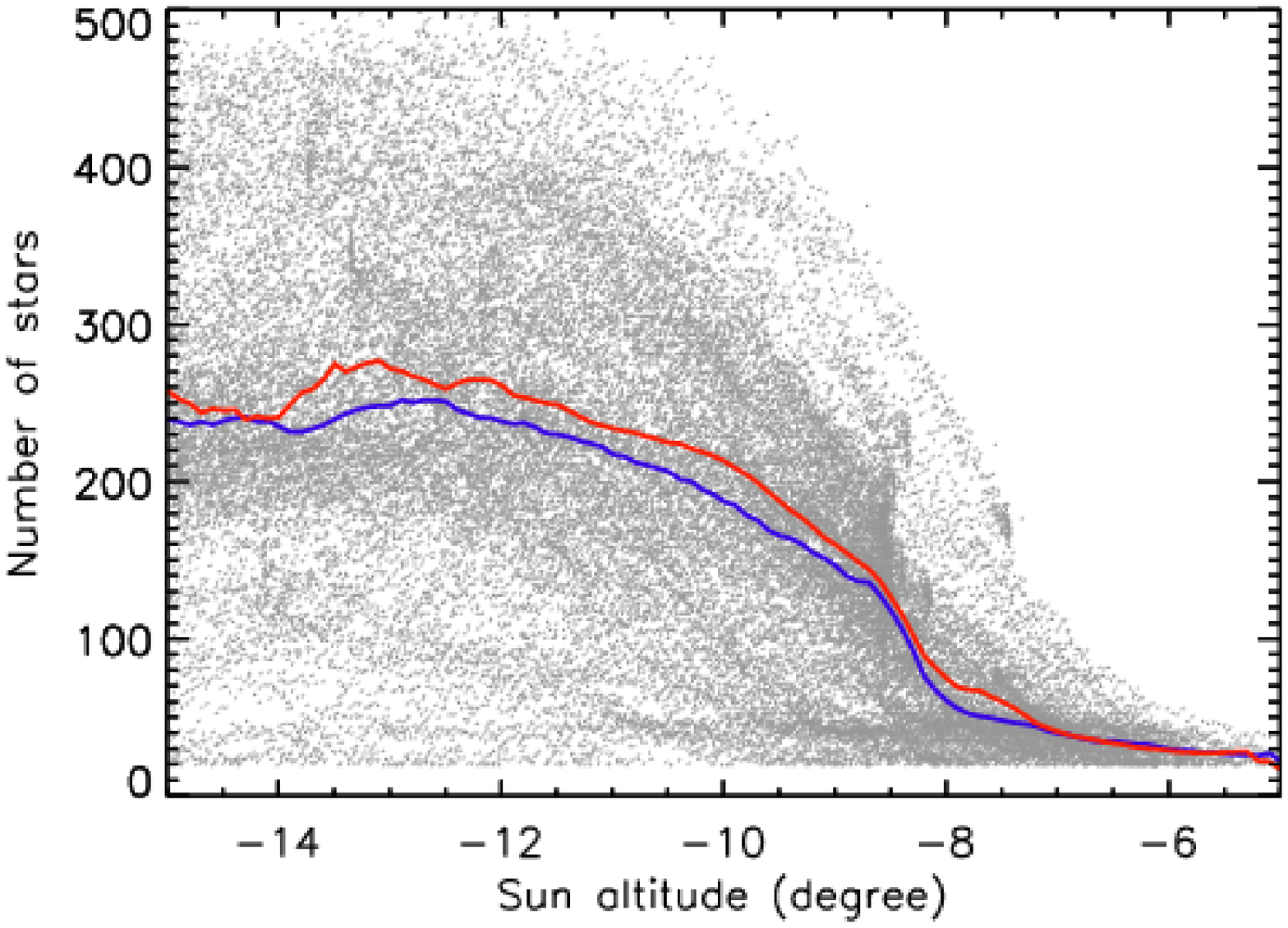}
\caption{Mean intensity of each image and number of stars as a function of the Sun altitude when it is between $-15\degree$~and~$-5\degree$. The bias level is around 2480~ADU and the intensity scale is the whole CCD dynamics. Points: individual measurements, blue and red: average during ascending and descending periods respectively. At $9\degree$ below the horizon the sky brightness is 16.6~mag\,/\,arcsec$^2$.}
\label{fig:Intenssun}
\end{figure}

The mean intensity of each image and the number of detected stars are plotted against the height of the Sun in figure~\ref{fig:Intenssun}. The fact that the sky intensity drops to an undetectable level when the Sun is below $-13\degree$ appears to be in line with the result from \citet{Moore2008} that the Dome~C sky may be darker than other sites. However, this conclusion is at most tentative due to the absence of a bias removal and dark sky magnitude determination. We notice that the R-band sky magnitude averaged over all observations for a Sun altitude of $-9\degree$ is $16.6\,\rm arcsec^{-2}$, very similar to that obtained close to the zenith for Paranal in the R-band, i.e. $16$ to $17\,\rm arcsec^{-2}$ \citep[][see their figure~6]{Patat2006}. 

\subsection{Influence of the Moon}

\begin{figure}[ht]
\centering
\includegraphics[width=9cm]{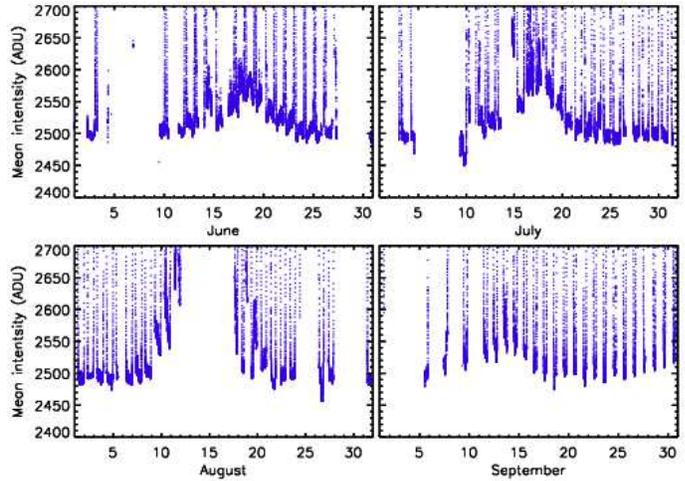}
\caption{Image intensity. The bias level is around 2480~ADU and the scale is 1~\% of the CCD dynamics. The sky background increases during 10~days around the full Moon \modif{up} to typically 18.1~mag\,/\,arcsec$^2$.}
\label{fig:Intensmonth_moon}
\end{figure}

The influence of the Moon is shown in figure~\ref{fig:Intensmonth_moon}. The Moon is full on June~18, July~18, August~16 and September~15. An increased sky background is clearly seen around these days, up to 80~ADU in June, 100~ADU in July, 500~ADU in August and 70~ADU in September. The full Moon in June and July corresponds to a good weather, without clouds, and the increase in intensity is low enough to allow \modif{photometric observations}. \modif{In contrast}, during the full Moon of August the weather was very bad with high temperature (up to $-$30\degree C), strong wind at ground level (up to 11~m/s), and a very cloudy sky. The very high background is thus interpreted as due to the reflection of moonlight by clouds. A typical increase of 80~ADU during the full Moon leads to a sky brightness of $\approx$ 18.1~mag\,/\,arcsec$^2$. As discussed in section~\ref{sec:calibration}, this magnitude estimate may change by a fraction of a magnitude with a precise bias subtraction. 



\subsection{Point Spread Function variations}

\begin{figure}[ht]
\centering
\includegraphics[width=9cm]{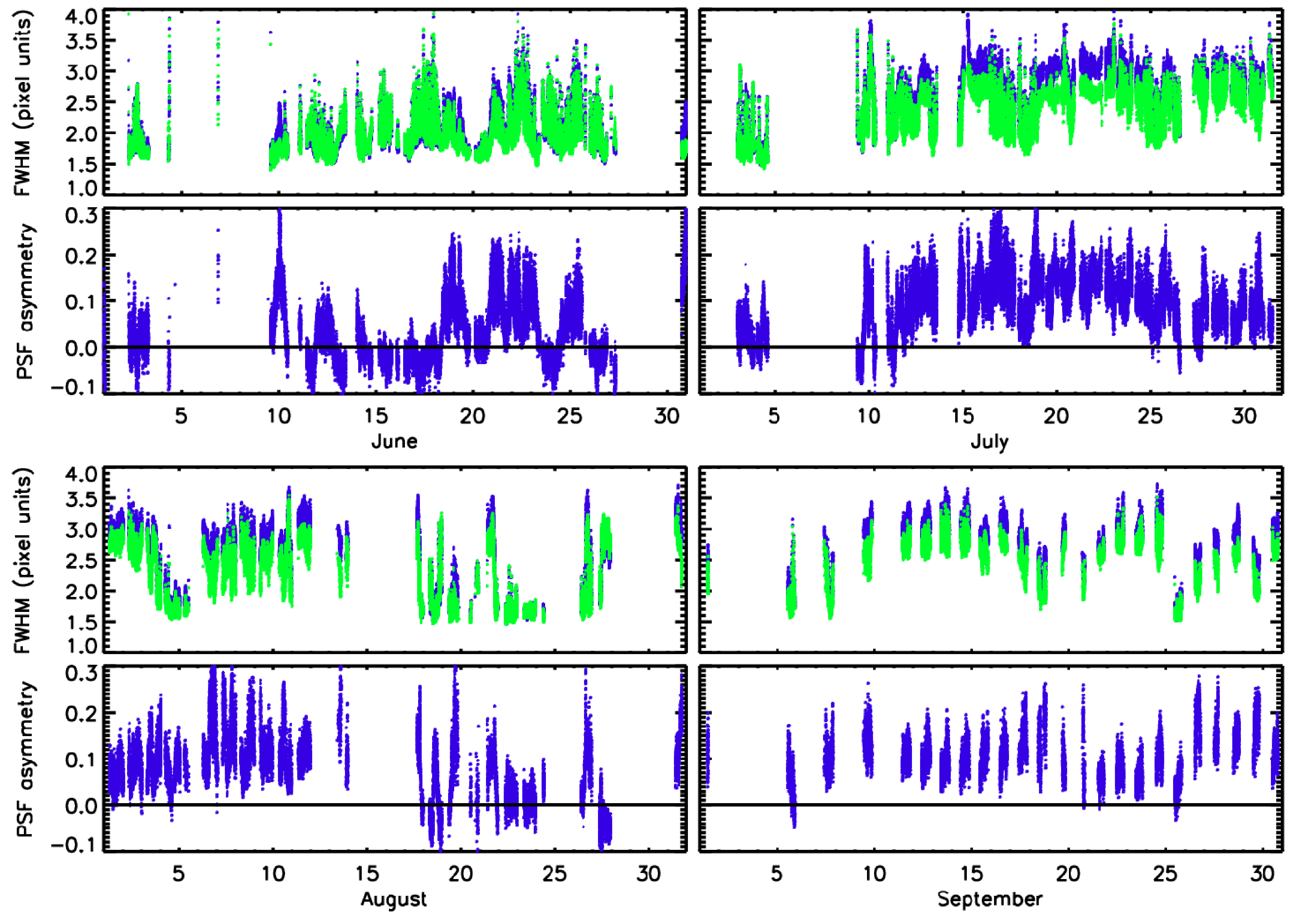}
\caption{\modif{Mean values of the size of the stars (FWHM) on the CCD in pixels (top panels) and their asymmetry (bottom panels) as a function of the observing day for the months of June (top left), July (top right), August (bottom left) and September (bottom right). In the top panels, the blue and green curves correspond to the values of the FWHM in the x and y directions, respectively. The mean FWHM values are obtained through a spatial average on the CCD.}}
\label{fig:PSFmonth}
\end{figure}

PSF variations are a crucial issue for photometry. We investigate here the PSF variations in the ASTEP South data. For each image, the 30 brightest stars are fitted with a gaussian PSF and their FWHM in both direction is calculated using DAOPHOT. \modif{The mean of the FWHM across the entire image is shown as a function of time in figure~\ref{fig:PSFmonth}. This mean FWHM varies between 1.5 and 3.5~pixels over the winter.}

\begin{figure}[ht]
\centering
\includegraphics[width=9cm]{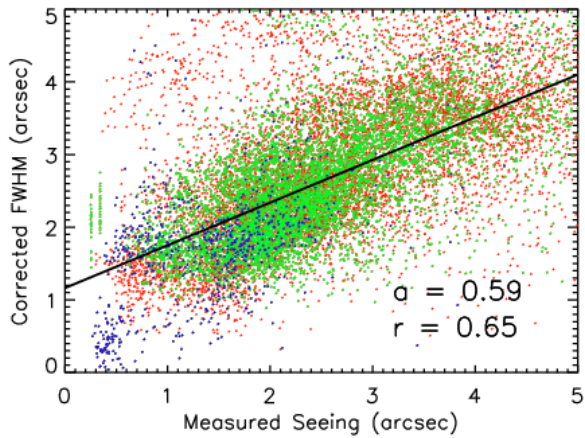}
\caption{Correlation between the PSF FWHM and \modif{independent} seeing measurements at Dome~C, for timescales smaller than one day (the FWHM is corrected for larger timescale variations). \modif{Direct seeing measurements from three Differential Image Motion Monitors are shown in blue, green and red.} A linear regression gives a slope $a = 0.59$ with a correlation coefficient $r = 0.65$. The PSFs are clearly affected by seeing variations at the ground level.}
\label{fig:seeing-psf}
\end{figure}

Two kinds of variations are present. First, PSF variations on a timescale smaller than one day are observed. We compare them to \modif{independent} seeing measurements at Dome~C provided by three dedicated Differential Image Motion Monitors (DIMM), two of them forming a Generalized Seeing Monitor (GSM) \modif{\citep[for a description of these instruments see][]{Ziad2008}}. \modif{In order to} consider only the PSF variations of period smaller than one day we \modif{subtract} to the FWHM the difference between the median FWHM and the median seeing \modif{for each day}. Figure~\ref{fig:seeing-psf} shows \modif{that on this day timescale the corrected FWHM and the seeing are clearly correlated: the PSF variations on short timescales are mostly due to seeing variations. As previously discussed,} the seeing at the ground level where ASTEP South is placed is rather poor: the median value in winter at 3 meters high reported by \citet{Aristidi2009} is 2.37 arcsec with stability periods of 10 to 30 minutes. This explains the \modif{short-term} variations of our PSFs.

On a timescale larger than one day the correlation is not true anymore. This shows that another cause of PSF variations is present. For this larger timescale, two regimes seem to be present, one with a PSF around 1.5 pixels and another with a PSF around 3.0 pixels. These variations are associated with an \modif{asymmetrical} deformation of the PSF. This suggests an instrumental cause of PSF variations such as astigmatism and decollimation. Indeed, temperature inhomogeneities in the thermal enclosure cause mechanical and optical deformations. Unfortunately these large timescale variations \modif{prevent us from estimating the seeing at the ground level directly from our photometric data.}




\subsection{Astrometry and pointing variations}

\begin{figure}[h]
\includegraphics[width=8cm]{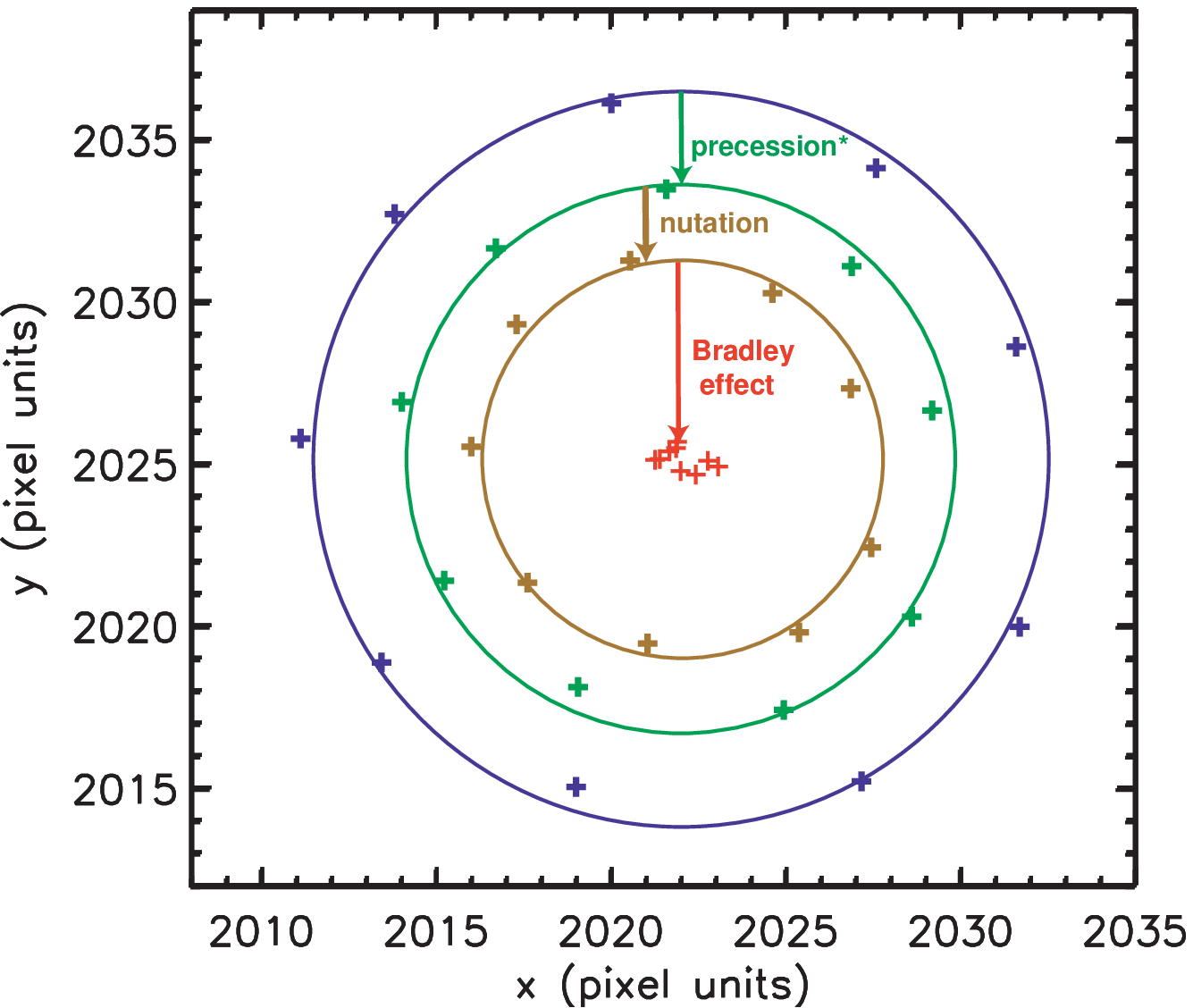}
\includegraphics[width=8cm]{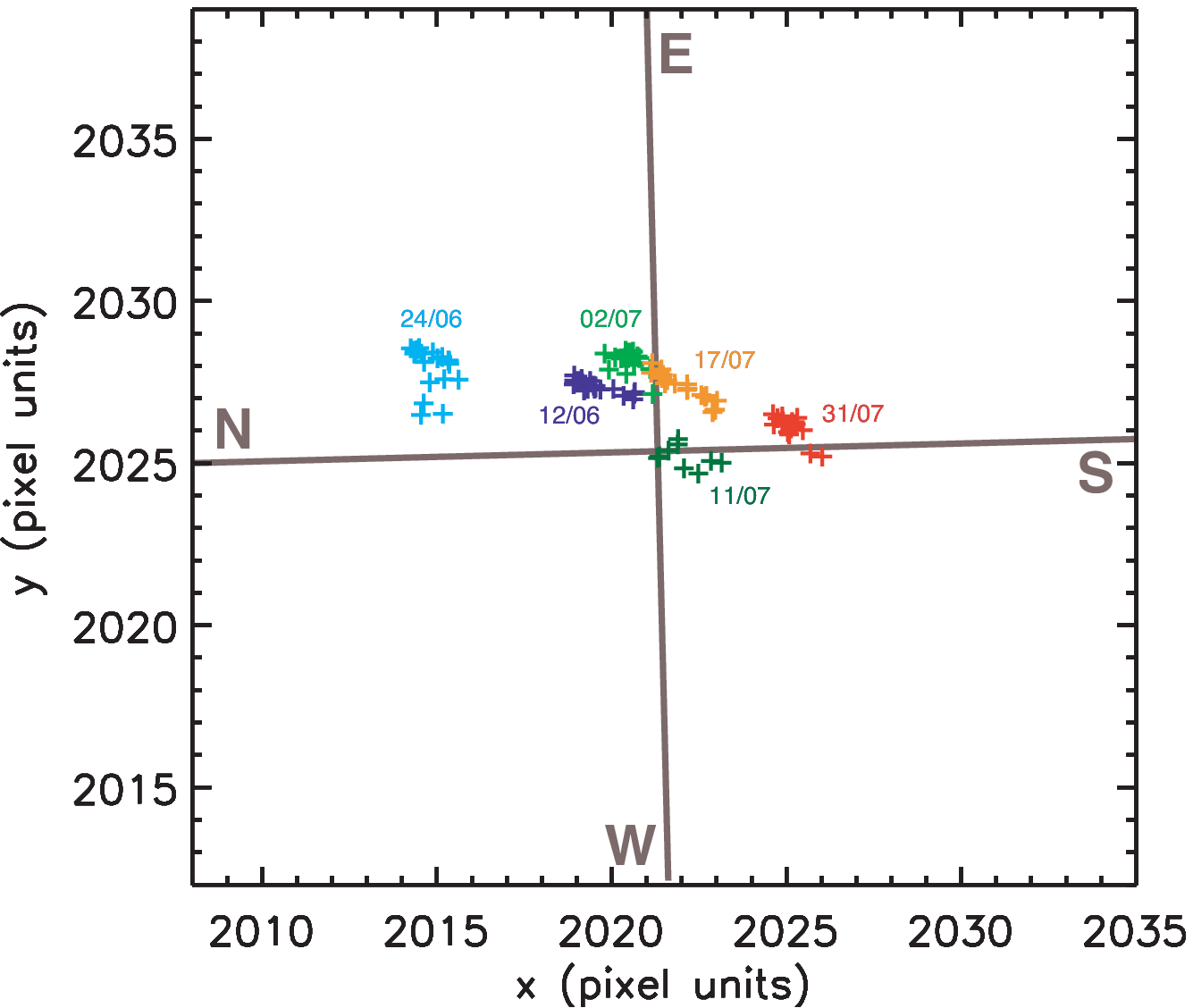}
\caption{Position of the pole on the CCD. Top: 9 images on July 11 after various astrometric corrections (blue: approximate correction of the precession of the equinox, green: improved correction of the precession of the equinox, brown: correction of the nutation, red: correction of the Bradley effect). Bottom: 7 days between June 12 and July 30, showing a drift over the winter (dark blue: June 12, light blue: June 24, green: July 2, yellow: July 11, orange: July 17, red: July 30).}
\label{fig:positionpole}
\end{figure}

Ideally the pointing should remain stable during all the winter, meaning that the South pole must stay at the same place on the CCD. The position of the South pole on the CCD is found on each image using a home-made \modif{field-matching} algorithm. The precision of this algorithm is typically 0.2 pixels. The results for a typical day and for all the winter are shown \modif{in} figure~\ref{fig:positionpole}. 

First we have a variation of this position with a period of one \modif{sidereal} day. This is due to an \modif{incomplete correction of astrometric effects}. Indeed, the star coordinates from the GSC2.2 catalog were corrected only for the precession of the equinox from the J2000 epoch to January 1, 2008. The remaining error on the star coordinates led to an error of 10 pixels (34 arcsec) in the determination of the position of the pole. We then corrected the GSC2.2 coordinates from the precession of the equinox using the real observation date, and from the nutation and the aberration of light (or Bradley effect). After these astrometric corrections the pole stays within 2 or 3 pixels during the day.\\
\indent Second, the pole is drifting during the winter. The amplitude is 10 pixels (34 arcsec) in 50 days, from June 12 to July 31. From the orientation of the CCD we find that this drift is oriented vertically towards the North. This may be due to mechanical deformations of the instrument, atmospheric changes or a motion of the ice under the instrument. In any case this effect is very small.

\subsection{Camera temperature variations}

\label{sec:Camera temperature variations}

\begin{figure}[b!]
\centering
\includegraphics[width=9cm]{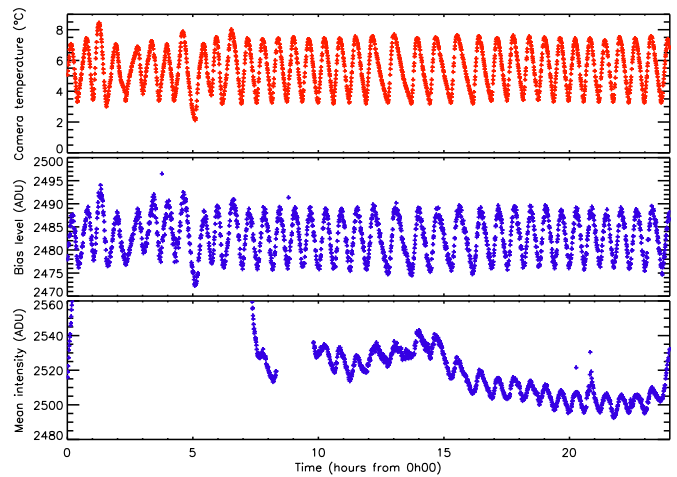}
\caption{Camera temperature, bias level and image intensity on July 11. The camera temperature varies between $+$4 and $+$8\degree C with a period of one hour and affects the bias level and the image intensity to about 10~ADU.}
\label{fig:Oscillations}
\end{figure}

\begin{figure}[b!]
\centering
\includegraphics[width=8cm]{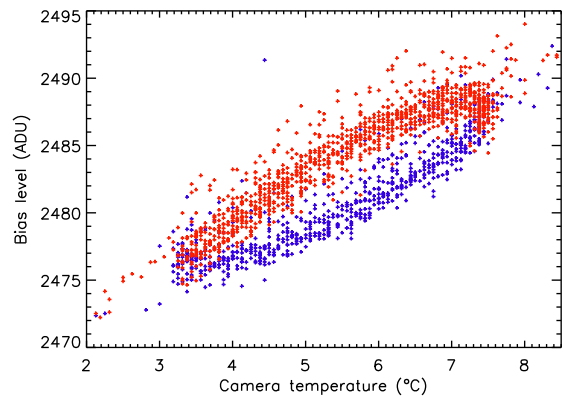}
\caption{Bias level against the camera temperature for July 11. The blue points correspond to an increasing temperature and the red points to a decreasing temperature, showing a hysteresis behavior.}
\label{fig:Tcam-bias}
\end{figure}

The CCD is cooled down to $-$35\degree C without any variations. \modif{In contrast} the electronic part of the camera oscillates between $+$4 and $+$8\degree C  with a one hour period (figure~\ref{fig:Oscillations}). A threshold effect explains these variations: the thermalized enclosure is not heated continuously but only when it passes below a threshold temperature. A direct consequence is seen on the bias images. The bias level oscillates with the same period and an amplitude of 10~ADU. The mean intensity of science images is affected in the same way. The bias level is plotted against the camera temperature in figure~\ref{fig:Tcam-bias} and shows a hysteresis behavior. For a given temperature, the bias level is lower if the temperature is increasing than if the temperature is decreasing. The hysteresis amplitude is around 5~ADU. An explanation can be that the temperature sensor is not \modif{exactly on} the electronics but \modif{is stuck on a camera wall which may be sensitive to temperature variations with a time lag compared to the electronics. It may also be due to the electronics and sensor having different thermal inertia.}

\section{Duty cycle}
\label{sec:duty cycle}

\begin{figure}[ht]
\centering
\includegraphics[width=9.4cm]{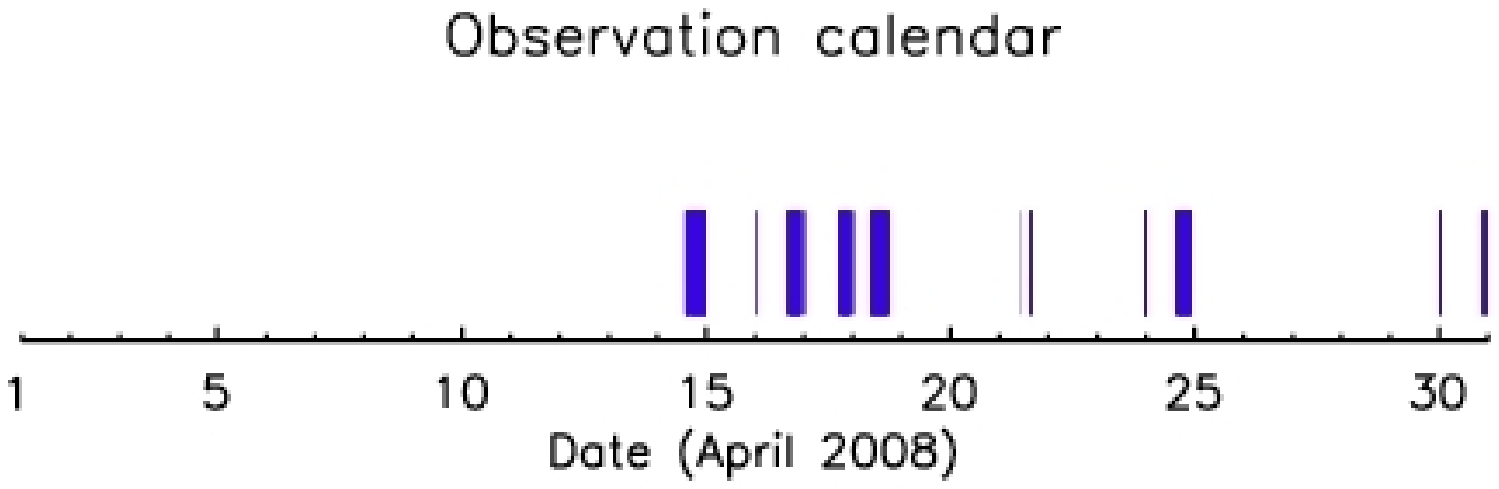}
\includegraphics[width=9.4cm]{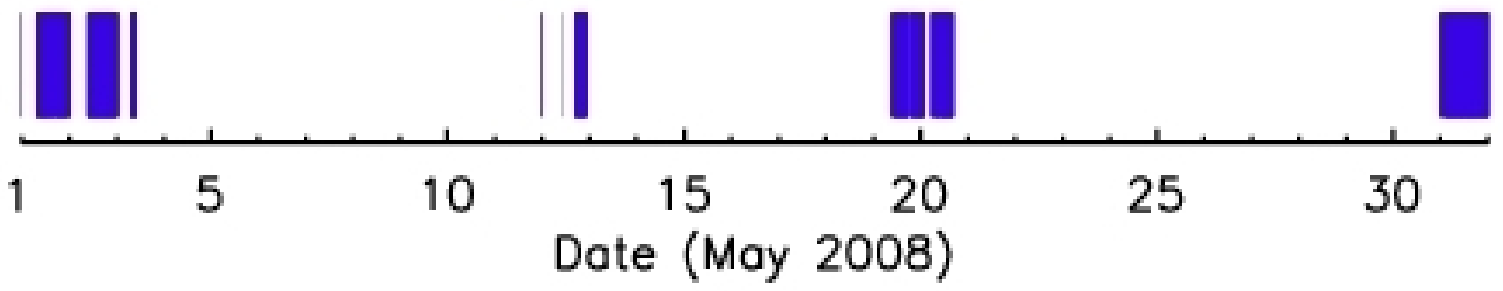}
\includegraphics[width=9.4cm]{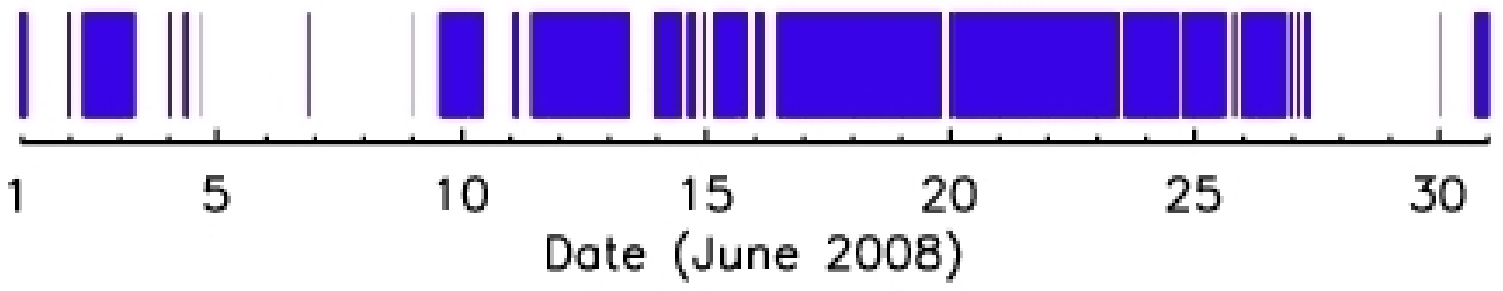}
\includegraphics[width=9.4cm]{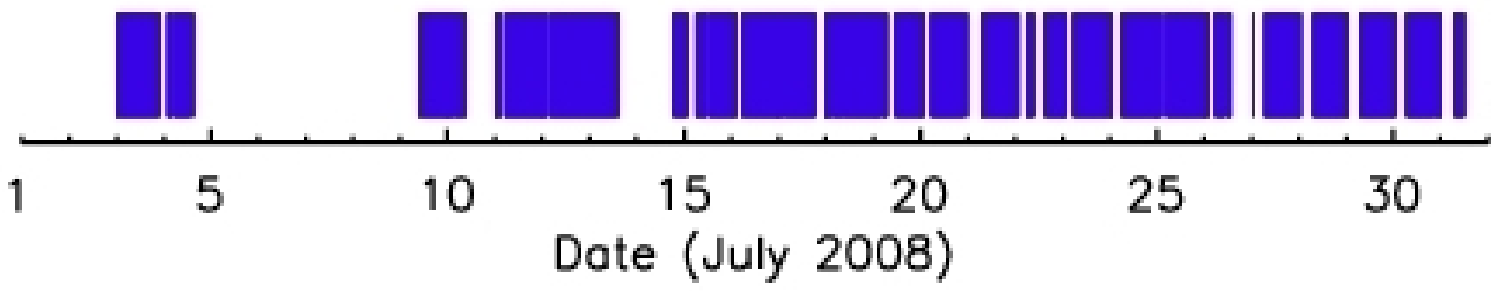}
\includegraphics[width=9.4cm]{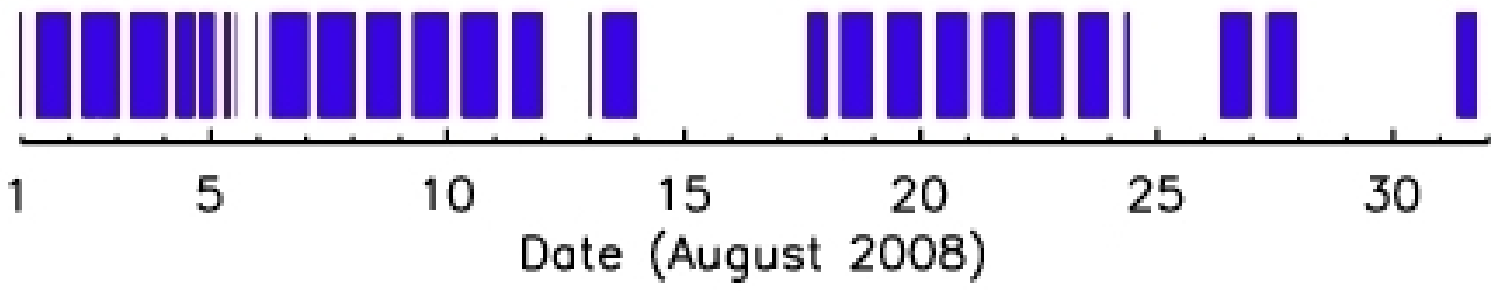}
\includegraphics[width=9.4cm]{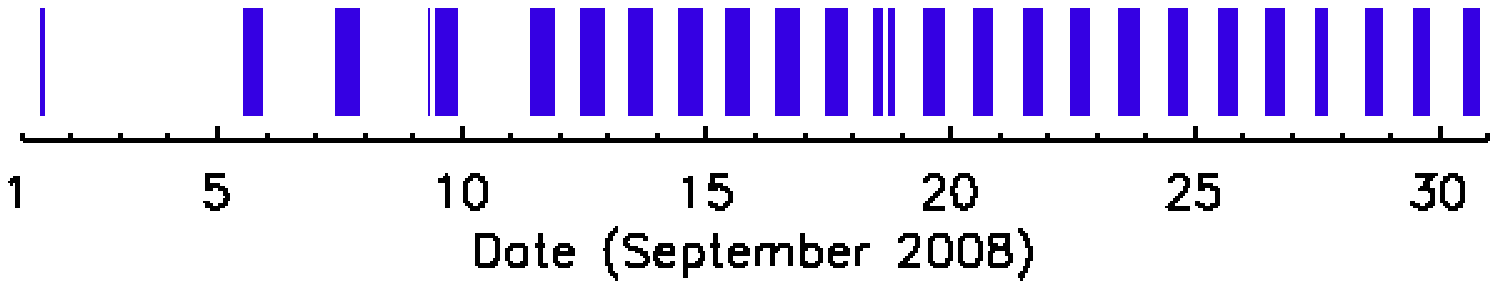}
\caption{ASTEP South observation calendar for the 2008 campaign. April and may were mainly dedicated to solving technical problems. Continuous observations started mid-June with very few interruptions until the end of the winter.}
\label{fig:ObsCalendar}
\end{figure}

A main objective of ASTEP South is to qualify the duty cycle for winter observations at Dome~C. The observation calendar for the whole 2008 campaign is shown \modif{in} figure~\ref{fig:ObsCalendar}. April and May were mainly devoted to setting up the instrument and software programs. Continuous observations started \modif{around} mid-June. Since then, very few interruptions \modif{occurred} and data were acquired until October. The effect of the Sun and \modif{of} the Moon has already been discussed in section~\ref{sec:data}. We present here some technical limitations to the duty cycle, and quantify the photometric quality of the Dome~C site for this campaign.

\subsection{Technical issues}

Technical issues encountered during this campaign limited the duty cycle. We show here typical issues \modif{that} instruments at Dome~C have to face with. We believe these can be mostly overcome with appropriate technical solutions.
\begin{itemize}  
\item First, the shutter did not close and got damaged at temperatures below 0\degree C. We had to change the shutter and build a special thermalization device to warm it.
\item \modif{In order to} install the camera again after changing the shutter, the thermalized box was opened and \modif{suddenly cooled by the ambient air at $\sim -60^\circ$C. As a result,} cables not made in teflon broke as well \modif{as} the camera USB connection. These \modif{had to be} replaced.
\item Outside instruments are affected by power cuts lasting \modif{for a} few minutes to \modif{a} few hours. The fraction of time lost for observations is negligible, however next instruments should be equipped with converters to avoid \modif{possible} damages.
\item The instrument is submitted to temperature gradients inside the thermal enclosure, and to the external temperature during power cuts or when opening the box. This leads to mechanical constraints resulting in decollimation and astigmatism.
\end{itemize}




\subsection{Weather conditions at Dome~C}

A first experiment to measure the winter clear sky fraction at Dome~C was made by \citet{Ashley2005b} with ICECAM, a CCD camera with a lens of $30\degree$ field of view. Every 2 hours from February to November 2001, ten images of the sky were taken and averaged. An analysis of all the images yielded a fraction of 74~\% of cloud-free time. An analysis by \citet{Mosser2007a} for the 2006 winter yielded an estimate of 92~\% of clear sky fraction by reporting several times a day the presence of clouds with the naked eye. \citet{Moore2008} derive a clear sky fraction of 79~\% for the winter 2006 from the Gattini instrument using the number of stars and the extinction across the images. In a previous work, we derived a clear sky fraction of 74~\% for the 2008 winter from the ASTEP South data, considering that the sky is clear if we have more than half of the stars detected on the best images \citep{Crouzet2010}. However this result is dependent on this ad hoc criterion. We reevaluate here this fraction by avoiding such an arbitrary limit.

\subsubsection{Method}

A new measurement of the clear sky fraction is made with ASTEP South using a method sensitive to thin clouds, based on the number of stars detected in the field. In order to do so, we need to evaluate the number of stars that should be detected on any given image if the weather was excellent. Our PSF size varies due to fluctuations of the seeing and of the instrument itself, and the background level also changes due to the presence of the Moon and the Sun. Since these are not directly related to weather, we need to derive how the number of detectable stars changes as a function of these parameters. (Note that thin clouds should affect the seeing in some way, however a posteriori examination of the data shows that this effect is small compared to the global attenuation due to clouds). 


\subsubsection{Identifying point sources with DAOPHOT }

The $1000\times 1000$ pixel sub-images contain up to 500 stars of varying magnitude. Our automatic procedure for finding point sources uses the FIND routine from DAOPHOT. Specifically, in this procedure a star is detected if the central height of the PSF is above the local background by a given number of standard deviations of that background. This threshold parameter $\alpha$ is chosen by the user\modif{. We choose $\alpha=5$.}

\subsubsection{A model to evaluate the expected number of stars}

The full width half maximum $\omega$ of our PSFs is typically 2.5 pixels. To evaluate whether a star is detected or not we compare the amplitude $A$ of the PSF to the noise in the central pixel. We consider two kind of noises: the photon noise from the sky background $N_{\rm sky}=\sqrt{F_{\rm sky}}$ and the read-out noise $N_{\rm ron}$\modif{. The noise} in the central pixel is \modif{hence} $ N= \sqrt{N_{\rm sky}^2+N_{\rm ron}^2} $. In order to obtain $A$ as a function of $\omega$ we consider a gaussian PSF. The amplitude $A$ of a gaussian is $ A = c F / \omega^2 $ with $F$ \modif{beeing} the total flux under the PSF and $ c \approx 0.88 $. The condition to detect a star $ A \ge \alpha N $ can thus be expressed as $ F \ge \alpha N \omega^2 / c $. The limit magnitude $m$ is therefore:
\begin{equation}
m = -2.5  \log( \alpha N \omega^2 / c ) + Z
\label{eq:maglim}
\end{equation}
with $Z$ being the photometric calibration constant. We have $N_{\rm ron} = 10$ electrons and set $\alpha = 5$ as we do in DAOPHOT, and since the instrument is not calibrated photometrically we \modif{use as} ad hoc constant $Z = 21.6$. As an example, typical values in our data are $F_{\rm sky} = 80$ electrons and $\omega = 2.5$ pixels. This \modif{yields} $m = 14.9$.

To derive the number of stars $N_*$ expected in a ${1000\times 1000}$ pixel sub-image from this limit magnitude we use a typical image taken under excellent weather conditions. We calculate the distribution in magnitude of the detected stars and fit it with a $3^{rd}$ order polynomial. For magnitudes larger than 14 the number of stars increases \modif{more slowly} because they are becoming too faint to be all detected\modif{. We therefore extend} the fitting function with a constant slope. \modif{The following relation provides our assumed} number of stars as a function of the limit magnitude:

\begin{equation}
  \log{N_*} = 
  \begin{cases}
    a_3 m^3 + a_2 m^2 + a_1 m + a_0 & \text{if } m \le 14 \\
    \log{N_{*14}} + 0.2\; (m-14) & \text{if } m > 14
  \end{cases}
  \label{eq:nbstars_mag}
\end{equation}

where $a_3=0.013$, $a_2=-0.664$, $a_1=11.326$, ${a_0=-61.567}$ and $N_{*14}$ is the number of stars detected for ${m=14}$. Equations (\ref{eq:maglim}) and (\ref{eq:nbstars_mag}) thus provide the number of stars that should be detected for a clear sky given a value of sky background and FWHM. In order to test the validity of this relation, we compare this to the maximum number of stars detected in our images for given values of FWHM and sky background. (By choosing the maximum number of stars, or more precisely the number of stars which is exceeded only 1~\% of the time, we ensure that we consider only images taken under excellent weather conditions). We find \modif{that} both agree with a standard error of 6.6~\% and a maximum error of 15~\%.

\subsubsection{Comparison to the measured number of stars}

\begin{figure}[ht]
\centering
\includegraphics[width=9.4cm]{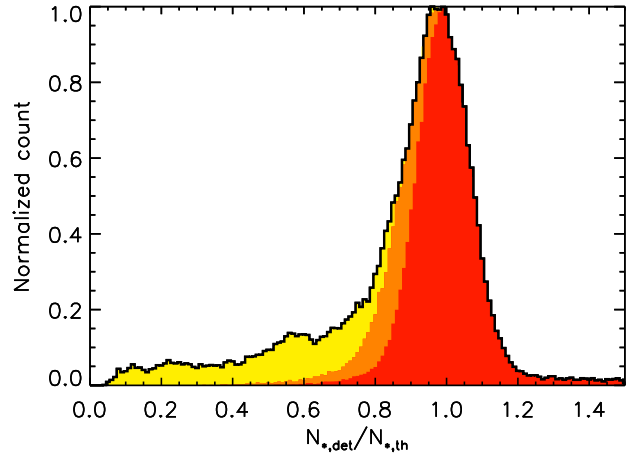}
\caption{Ratio between the measured and the expected number of stars (black line). Theoretical histograms taking into account low and high measurement errors allow to identify 3 parts in the data: excellent weather in red, uncertain weather in orange and veiled weather in yellow.}
\label{fig:weather_histogram}
\end{figure}

We use this model to compare the measured to the expected number of stars for each data point, given its FWHM and sky background. The resulting distribution in figure~\ref{fig:weather_histogram} shows \modif{two features}: a main peak centered around 1 with most of the points, meaning that for these points the weather is excellent, and a tail for which the measured number of stars is smaller than expected, corresponding to veiled weather.

The spread around the main peak of the histogram is partly due to measurement errors in the sky background and FWHM. The limit between this natural spread and the veiled weather data points must be defined to calculate the weather statistics. To do this we use a large set of sky background and FWHM values coherent with our measurements, and calculate the expected number of stars for each point ${(F_{\rm sky},\omega)}$ according to our model. We then add some random errors to this set of values and calculate again the expected number of stars. For each point ${(F_{\rm sky},\omega)}$ we compare the expected number of stars obtained with the added errors to the one without errors. This \modif{yields} a theoretical distribution of the number of stars \modif{that accounts} for measurement errors.

The sky background value is typically around 40~ADU. We estimate the error to be around 10~\% of this value, i.e. only 4~ADU. This error is not dominant and the spread of the theoretical distribution is mainly due to the error on the FWHM. The FWHM is typically between 2 and 3 pixels, and we suppose again an error of 10~\%, i.e. 0.25 pixels. However the shape of the theoretical distribution does not fit perfectly to the data. Instead we find that we can fit the two sides of the main peak in figure~\ref{fig:weather_histogram} with two different errors on the FWHM, corresponding to a low and a high measurement error, respectively 0.17 and 0.25 pixels.

The data points fitting into the theoretical distribution with the low error are considered as excellent weather (red part in figure~\ref{fig:weather_histogram}). Those between the low and high error distributions can be either due to a large measurement error or to veiled weather, thus they are considered as uncertain weather (orange part). The data points outside the low error distribution correspond to veiled weather (yellow part).

\subsubsection{Weather statistics for the winter 2008}
\label{sec:Weather statistics for the 2008 winter}

\begin{figure}[ht]
\centering
\includegraphics[width=9.4cm]{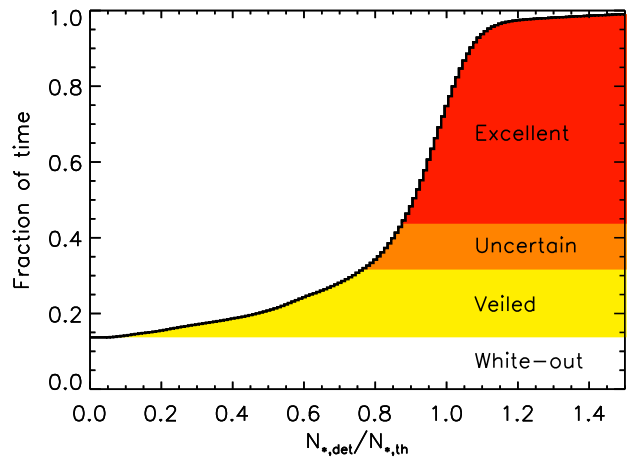}
\caption{Cumulative histogram of the ratio between the measured and the expected number of stars. The data are distributed between excellent, uncertain and veiled weather. The white part is the fraction of time during which the acquisitions where stopped because of very bad weather (white-out).}
\label{fig:weather_cumulative_histogram}
\end{figure}

This analysis gives a fraction of time between 65.2~\% and 79.2~\% with excellent weather and between 20.8~\% and 34.8~\% of veiled weather. Only the periods with data when at least few stars are visible are considered here, excluding in particular the white-out periods. During the winter the acquisitions were stopped during 13.7~\% of time because of very bad weather, so the previous numbers must be multiplied by ${1-0.137 = 0.863}$. The weather statistics for the 2008 winter at Dome~C are therefore: between 56.3~\% and 68.4~\% of excellent weather, 17.9 to 30~\% of veiled weather during which stars are still visible and 13.7~\% of bad weather (figure~\ref{fig:weather_cumulative_histogram}). For comparison the fraction of photometric weather during night-time is 62~\% at La Silla and 75~\% at Paranal as provided by M.Sarazin\footnote{\tt http://www.eso.org/gen-fac/pubs/astclim/paranal/\\clouds/statcloud.gif} \citep[see also][]{Ardeberg1990}, whereas it is only 45~\% at Mauna Kea \citep[][and references therein]{Ortolani2003} though more recent results report 56~\% \citep{Steinbring2009}. Dome~C is therefore very competitive compared to other observing sites.


\subsubsection{Validation}

\begin{figure}[ht]
\centering
\includegraphics[width=9cm]{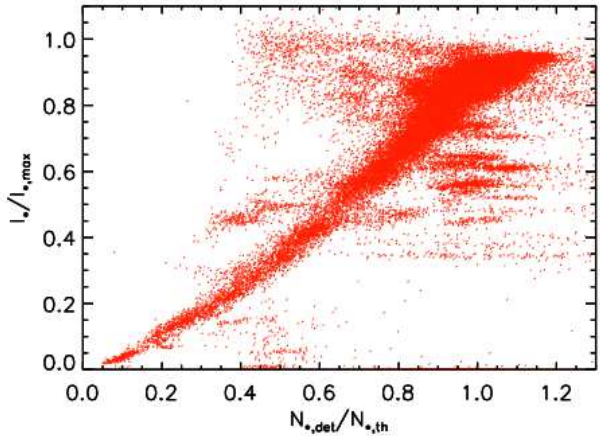}
\caption{Normalized intensity of \modif{a selected sample of nine stars} as a function of the ratio between the measured and the expected number of stars \modif{based on theoretical estimates (see text)}.}
\label{fig:nbstars_starintensity}
\end{figure}

To validate the method, we compare the number of stars to the intensity received from the stars, as both should be smaller if clouds are present. We measure the intensity received from nine stars of magnitude 8 to 12 and sum them after dividing each star by its median value. We then normalize this sum by the maximum value, more precisely by the mean of the 1~\% highest values. We use only the periods with a moderate sky background i.e. when the Sun is below $-13\degree$ and excluding the full Moon periods (the result is however very similar using all data points). Figure~\ref{fig:nbstars_starintensity} shows the normalized star intensity as a function of the ratio between the measured and the expected number of stars. As expected, both parameters are directly correlated, thus validating the method. We further note that for data points taken under excellent or intermediate weather conditions, i.e. when $N_*/N_{*th} > 0.7$, the normalized intensity received from the stars is $0.77 \pm 0.22$. Both parameters are thus good indicators of the cloud cover in the field of view.

\subsection{Auroras}

Aurorae were first feared to be a source of contamination for long-term photometric data. However, it is to be noted that Concordia is a favorable site in that respect: auroras occur mainly in the auroral zone, a \modif{ring-shaped} region with a radius of approximately 2500 km around the magnetic pole. The Concordia site is located well within this ring, only 1300 km from the South magnetic pole, and thus has a much lower auroral activity than other sites (e.g. Dome~A).

During the winter-over 2008, the Concordia staff reported 2 auroras on July 30 and 31. On July 30, a careful examination of the ASTEP South data indicates a possibility of auroral contamination between 14:12 and 14:24 UTC, but it cannot be distinguished from thin clouds. The July 31 data were unfortunately lost during the copy, probably because of a hard drive glitch (the only instance of that occurring) so that we could not attempt to check the images for that day.

In any case, the ASTEP South 2008 data were not contaminated by auroras, confirming the low contribution of auroras to the sky brightness as suggested by \citet{Dempsey2005}. It will be interesting to see whether it remains true when progressively moving towards a maximum solar activity in 2012.

\subsection{Observing time and photometric quality of Dome~C}

\begin{figure}[ht]
\centering
\includegraphics[width=8.5cm]{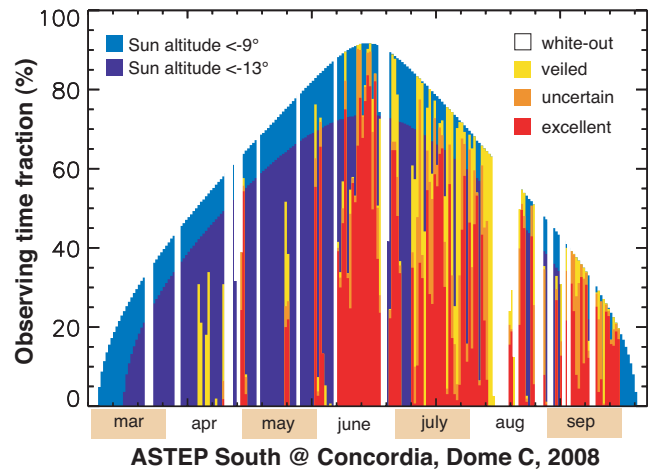}
\caption{\modif{Daily observing time fraction for ASTEP South in 2008 as a function of the observation period. The light blue and dark blue regions indicate the fraction of time for which the Sun is lower than $-9^\circ$ and $-13^\circ$ below the horizon, respectively. Periods of excellent, uncertain and veiled weather as observed by ASTEP South are indicated in red, orange and yellow, respectively. White areas correspond to periods during which observations were not possible, either because of the Sun altitude or because of bad weather.}}
\label{fig:DutyCycle}
\end{figure}

The duty cycle for the 2008 campaign of ASTEP South is represented in figure~\ref{fig:DutyCycle}. The limit due to the Sun, the observing time and the excellent and intermediate weather fractions are shown for each day, as well as the white-out periods. We acquired 1592 hours of data with ASTEP South on a single field during the 2008 campaign. From the previous analysis we have 1034 hours with excellent or uncertain weather. As a comparison, simulations based on the method described in \citet{Rauer2008} show that the time usable for photometry in one year at La Silla for the field with the best observability is typically 820 hours per year (see section \ref{sec:Planet detection probability} for more details). Moreover, the white-out periods at Dome~C last typically from one to \modif{a} few days, allowing \modif{extended periods of} continuous observations between them. For example we observed every day during one month between July 9 and August 8. Considering the excellent and uncertain weather and the hours lost because of the Sun, the fraction of time usable for photometry for this one month period is 52~\%. In La Silla the fraction of time usable for photometry for all one month periods between 1991 and 1999 has a mean of 27~\% with a maximum of 45~\% in April 1997 (from the La Silla weather statistics\footnote{\tt http://www.ls.eso.org/sci/facilities/lasilla/\\astclim/weather/tablemwr.html} multiplied by the night-time fraction). This shows the very high potential of Dome~C for \modif{continuous} observations during the Antarctic winter.



\section{Planet detection probability}

\label{sec:Planet detection probability}

\modif{As shown by \citet{Pont2005a}}, the high phase coverage at Dome~C should improve the efficiency of a transit survey. \modif{Here we} investigate the potential of ASTEP South for transit detection using CoRoTlux and compare the Dome~C and La Silla observing sites. CoRoTlux performs statistical simulations of transit events for a survey given the star distribution in the field of view, the instrumental parameters and the observation windows \citep{Fressin2007a,Fressin2009}. In all simulations, the star distribution is the one of the South pole field (see section \ref{sec:The South pole field}). We use the GSC2.2 catalog for stars from magnitude 10 to 14.5 completed with a distribution from the Besan\c{c}on model up to magnitude 18 (after scaling it to the GSC2.2 catalog for low magnitude stars). The target stars range from magnitude 10 to 15 and the background stars from magnitude 15 to 18. The instrumental parameters are always those of ASTEP South. We perform three simulations corresponding to three survey configurations differing only by their observation windows.

The first set of observation windows \modif{that} we have used in our simulations corresponds to the periods during which ASTEP South actually ran in excellent or uncertain weather conditions in 2008, i.e. the red and orange parts of the duty cycle in figure~\ref{fig:DutyCycle}. \modif{This provides us with the potential yield of the 2008 campaign in terms of detections of transiting planets}.

We also want to compare Dome~C and La Silla. In that purpose we consider an ideal campaign \modif{for an ASTEP~South-like instrument for which the observation windows are determined only from the altitude of the Sun and weather statistics at that site}. For Dome~C, we apply the weather statistics presented in section \ref{sec:Weather statistics for the 2008 winter} \modif{to an entire winter season in order to generate} the second set of observation windows. For this second simulation, we incorporate over the Sun limited duty cycle 13.7~\% of white-out periods and 17.9~\% of randomly distributed cloudy periods lasting less than one day.

\begin{figure}[ht]
\centering
\includegraphics[width=8.5cm]{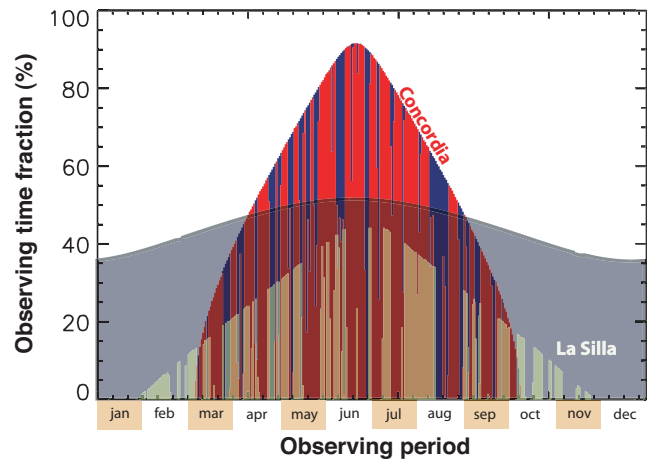}
\caption{\modif{Observing time fraction as a function of observing period at Concordia and La Silla. The blue and grey envelopes indicate values obtained by imposing a Sun altitude lower than $-9^\circ$ below the horizon, for Concordia and La Silla, respectively. The red histogram is an example of a generated window function for Concordia using the weather statistics obtained from ASTEP South. The green histogram is generated for the field which is observable the longest with a high-enough airmass at La Silla and using the 1987-2007 weather statistics of the site.}}
\label{fig:weathersun_dutycycle}
\end{figure}

\modif{For La Silla, we generate a third set of observation windows using the monthly weather statistics acquired from 1987 to 2007 \footnote{\tt http://www.eso.org/gen-fac/pubs/astclim/paranal/\\clouds/statcloud.lis}.} The weather statistics for each month is taken as the mean of the photometric fraction for this given month over all years. \modif{At La Silla, one cannot simply stare at the South pole field continuously because it is low over the horizon. A best pointing can be found that maximizes the observation time as a combination of weather statistics, night-time and airmass} \citep[for a complete description of the method see][]{Rauer2008}. The resulting field with the best observability is centered on RA$\,=\,$18h30' and DE$\,=-$58\degree 54'. For consistency with the other simulations we \modif{use the same stellar population as for the South pole field, and consider that photometric observations are possible when the Sun is less than $-9\degree$ below the horizon}. The resulting duty cycles for weather and Sun limited observations of a single field over one year are shown in figure~\ref{fig:weathersun_dutycycle} for both Dome~C and La Silla. The total observing time is typically 2240 hours for Dome~C and 820 hours for La Silla.

\begin{figure}[ht]
\centering
\includegraphics[width=9cm]{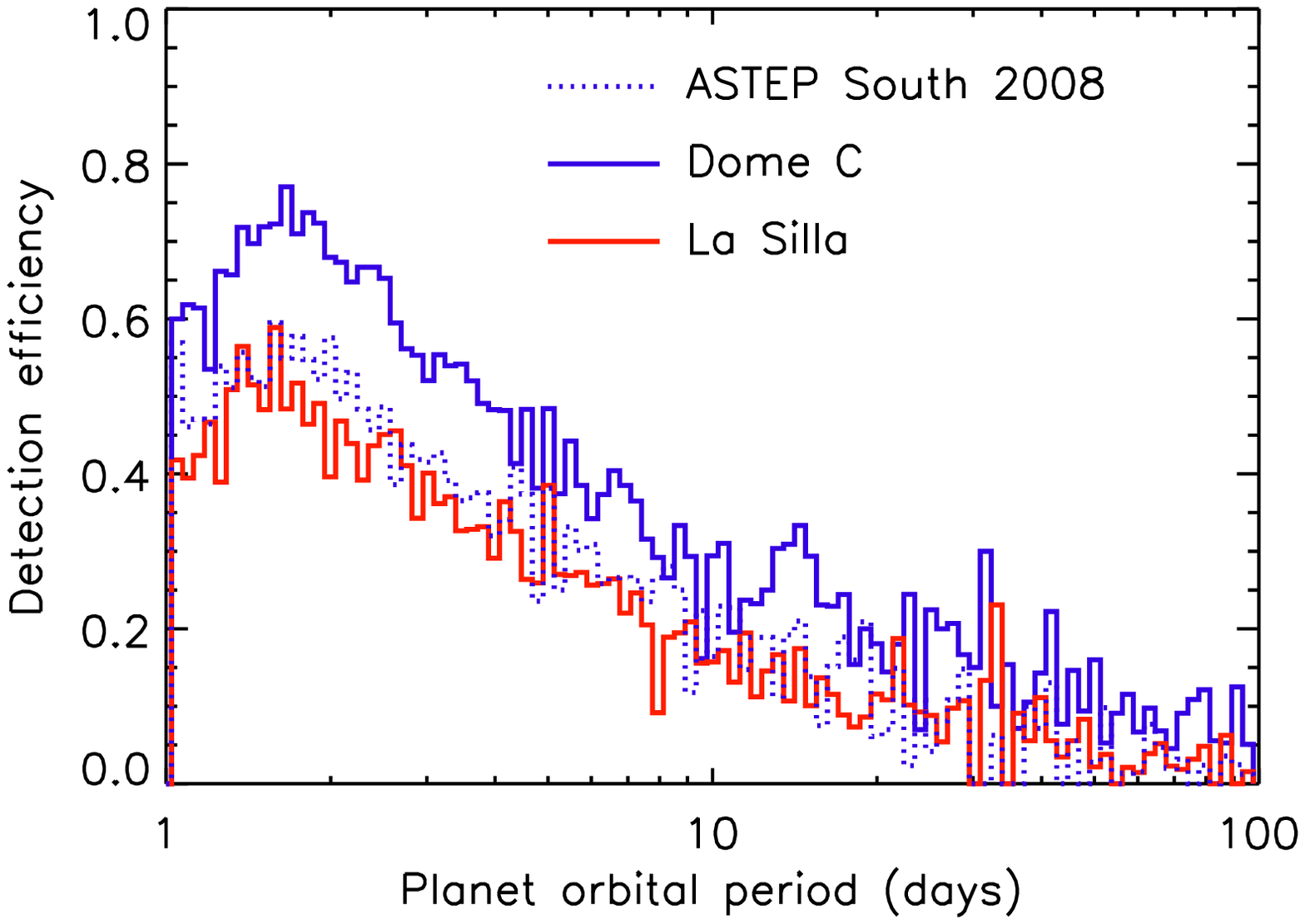}
\includegraphics[width=9cm]{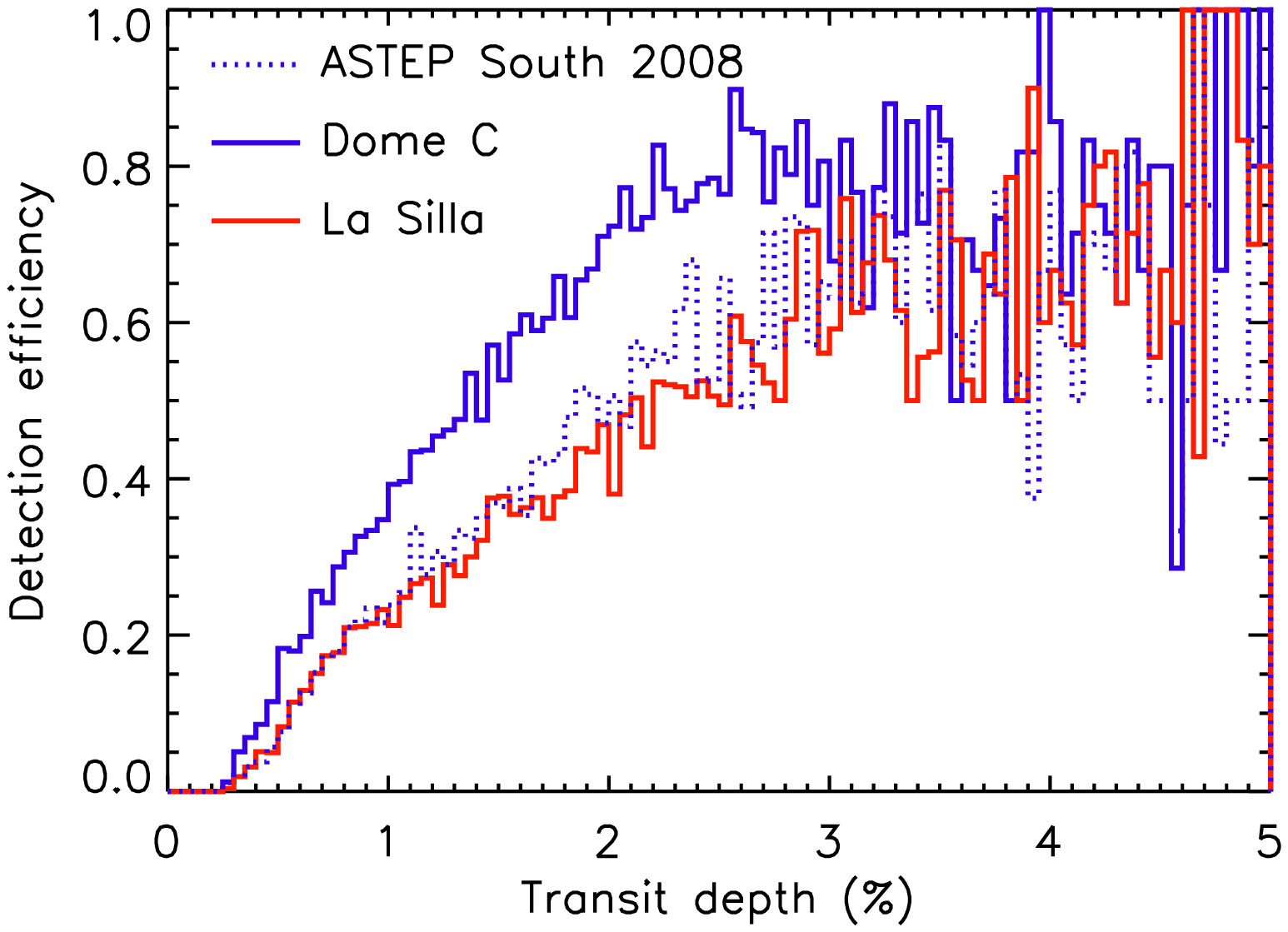}
\caption{\modif{Calculated efficiency of detection of transiting giant planets for a single field as observed by an ASTEP-South like survey during a full season, and as a function of the orbital period (top) and transiting depth (bottom). {\it Dotted blue lines}: Detection efficiency for the ASTEP South 2008 campaign. {\it Plain blue line}: Detection efficiency for a full winter at Dome C for a circumpolar field limited only by the weather statistics and the constraint of a Sun altitude below $-9^\circ$. {\it Plain red line}: Same as before, but for a survey at La Silla and the field with the best observability over the year (see Fig. 17 and text).}}
\label{fig:corotlux}
\end{figure}

A large number of runs \modif{($\sim 3000$) are performed} for each survey configuration \modif{in order} to have a significant statistic. The results of the simulations provide the number of detectable planets. We assume that only transiting planets with a signal to noise ratio higher than 10 are detectable. This yields 1.08 planets for the ASTEP South 2008 campaign (i.e. 3244 planets over 3000 runs), 1.62 planets for a whole winter at Dome~C, and 1.04 planet for La Silla. These numbers are low because ASTEP South is a small instrument, however the number of planets is notably higher for a survey from Dome~C than from La Silla. \modif{The resulting} planet detection efficiency \modif{is} shown in figure~\ref{fig:corotlux}. The detection efficiency is defined as the number of detectable planets divided by the total number of simulated planets. In spite of technical problems at the beginning of the winter, the detection efficiency for the ASTEP South 2008 campaign is equivalent to the one obtained for one year at La Silla. When comparing an instrument that would run for the entire observing season, the detection efficiency is found to be significantly higher at Dome~C than at La Silla both in terms of planet orbital period and transit depth. For example we have an efficiency of 69~\% at Dome~C vs 45~\% at La Silla for a 2-day period giant planet, and 76~\% at Dome~C vs 45~\% at La Silla for a 2~\% transit depth. The detection efficiency decreases for planets with longer orbital periods, but is even more favorable to Dome~C relatively to La Silla. On the other hand, it is true that a mid-latitude site offers more available targets. However, we believe that this shows the high potential of Dome~C for future planet discoveries. 


\section{Conclusion}

\modif{ASTEP South, the first phase of the ASTEP project observed 1592 hours of data. Night-time photometric observations started in a nearly continuous way around mid-June and proceeded to the end of September, when the sky became too bright even at midnight local-time. Our preliminary analysis showed that the Sun affects our photometric measurements when it is at an altitude higher than $-13^\circ$ below the horizon. The sky brightness at dusk and dawn appears to vary quite significantly from one day to the other, but its mean is very similar to results obtained close to the zenith at Paranal (an R-band sky-magnitude $\text{R}=16.6\,\rm arcsec^{-2}$ for a Sun altitude of $-9^\circ$). The full Moon yields a sky brightness of $\text{R}\approx 18.1\,\rm arcsec^{-2}$. Apart from one possible instance lasting only 12 minutes, auroras had no noticeable impact on the data.}

\modif{An identification of the stars in the field allowed us to retrieve the precise location of the celestial South pole on the images and show that the pointing direction is stable within 10 arcseconds on a daily timescale for a drift of only 34 arcseconds in 50 days. On the basis of the number of identified stars and of a model to account for PSF variations and sky brightness, we retrieved the weather statistics for the 2008 winter: between 56.3~\% and 68.4~\% of excellent weather, 17.9~\% to 30~\% of veiled weather (when the probable presence of thin clouds implies a lower number of detected stars) and 13.7~\% of bad weather.}

\modif{An analysis of the yield of transit surveys with our weather statistics at Dome C compared to those at La Silla showed that the efficiency to detect transiting planets in one given field is significantly higher at Dome C (69~\% vs. 45~\% for 2-day period giant planets with an ASTEP South-like instrument in one season). The prospects for the detection and characterization of exoplanets from Dome C are therefore very good. Future work will be focused on a detailed analysis of the full ASTEP South images. The second phase of the project includes the installation of ASTEP 400, a dedicated automated 40-cm telescope at Concordia and its operation in 2010.}

\section{Acknowledgements}

The ASTEP project is funded by the Agence Nationale de la Recherche (ANR), the Institut National des Sciences de l'Univers (INSU), the Programme National de Plan\'etologie (PNP), and the Plan Pluri-Formation OPERA between the Observatoire de la C\^ote d'Azur and the Universit\'e de Nice-Sophia Antipolis. The entire logistics at Concordia is handled by the French Institut Paul-Emile Victor (IPEV) and the Italian Programma Nazionale di Ricerche in Antartide (PNRA). The research grant for N.~Crouzet is supplied by the R\'egion Provence Alpes C\^ote d'Azur and the Observatoire de la C\^ote d'Azur. We wish to further thank E. Fossat for his helpful remarks on the manuscript. See http://fizeau.unice.fr/astep/ for more information about the ASTEP project.

\bibliography{biblio/ASTEPSud}

\end{document}